\documentclass[prd,amsmath,amssymb]{revtex4} 
\usepackage{graphicx} 
\usepackage{amsmath}
\usepackage{amsfonts,amsbsy}
\usepackage{amssymb}

\def\lsim{ \,\, \vcenter{\hbox{$\buildrel{\displaystyle <}\over\sim$}}
 \,\,}
\def\U{{\cal U}}
\def\be{\begin{equation}}
\def\ee{\end{equation}}
\def\bea{\begin{eqnarray}}
\def\eea{\end{eqnarray}}

\begin{document}

\title{\bf Particle production in high-energy collisions beyond the
  shockwave limit}

\author{Tolga Altinoluk$^a$ and Adrian Dumitru$^{b,c}$}
\affiliation{
$^a$ 
Departamento de F\'isica de Part\'iculas and IGFAE,
Universidade de Santiago de Compostela,
E-15706 Santiago de Compostela,
Galicia-Spain\\
$^b$ Department of Natural Sciences, Baruch College, CUNY,
17 Lexington Avenue, New York, NY 10010, USA\\
$^c$ The Graduate School and University Center, The City
  University of New York, 365 Fifth Avenue, New York, NY 10016, USA
}

\begin{abstract}
We compute next to eikonal (NE) and next to next to eikonal (NNE)
corrections to the Lipatov vertex due to a finite target
thickness. These arise from electric field insertions into the eikonal
Wilson lines. We then derive a $k_T$-factorization formula for single
inclusive gluon production at NNE accuracy and find that nuclear
effects are absent. We also analyze NNE corrections to two-gluon
production where some of the contributions are found to exhibit
corrections proportional to $A^{2/3}$.
\end{abstract}

\maketitle

Production of particles with moderately high transverse momentum in
high-energy hadronic collisions probes the gluon fields of the
projectile or target at small light-cone momentum
fractions~\cite{KovchegovLevinBook}. The field (in light-cone gauge,
$A^+=0$) in the forward light cone of a collision of two infinitely
thin charge sheets (shock waves) is given
by~\cite{Kovner:1995ja,Kovchegov:1997ke,Blaizot:2004wu}
\be \label{eq:Field}
p^2 A^{i,a}(p) = \left(T^a\right)_{bc} \, g^3 \int dz_1^- dz_2^+ \int
    \frac{d^2k}{(2\pi)^2} L^i(p,k)\,
    \frac{\rho^b_1(z_1^-,k)\, \rho^c_2(z_2^+,p-k)}{k^2\, (p-k)^2}~.
\ee
Here $p$ is the transverse momentum of the produced gluon and
$L^i(p,k)$ is the Lipatov vertex~\cite{Lipatov},
\be \label{eq:EikonalL}
L_i(p,k)\, L^*_i(p,q) = \frac{4}{p^2}\left[\delta^{ij}\delta^{lm} + 
\epsilon^{ij}\epsilon^{lm}\right] \, k^i (p-k)^j\, q^l (p-q)^m~.
\ee
In eq.~(\ref{eq:Field}) $\rho_{1,2}$ denote the random color
charge densities of projectile and target, respectively, which will be
averaged over. The equation is valid to leading order in both color
charge densities; a generalization to all orders in $\rho_{2}$ was
given in ref.~\cite{Dumitru:2001ux}. 

Eq.~(\ref{eq:EikonalL}) applies in the shockwave limit where the
projectile charges propagate on eikonal trajectories through the field
generated coherently by all valence charges in the
target. Ref.~\cite{Blaizot:2004wu}, for example, offers a very clear
discussion. However, at finite energies the non-zero thickness
$\ell^+$ of the target should be taken into account when $p^2 \,
\ell^+/p^+\sim p\, \ell^+ e^{-y}$ is not negligible. This is the case,
in particular, for heavy-ion targets since $\ell^+ \sim A^{1/3}$. We
should emphasize that our focus here is not on finite-$x$ corrections
to the small-$x$ evolution of the unintegrated gluon
distribution. Such evolution equations for some specific gluon
distributions have been derived in ref.~\cite{Balitsky:2015qba} to
order $(\ell^+/p^+)^0$.  Furthermore, kinematic finite energy
corrections not proportional to the target thickness have been derived
by Babansky and Balitsky~\cite{Babansky:2002my}; they find that such
corrections are important for dipole-dipole scattering at rapidity
$\lsim 5$.  Rather, here we consider corrections to the particle
production vertex beyond the shockwave approximation for the {\em
  valence} charges; this is a nuclear effect proportional to $A^{1/3}$
and should be relevant in particular for a heavy-ion target.

The gluon production cross section then involves one or more electric
field insertions into the eikonal Wilson
lines~\cite{Altinoluk:2014oxa,Altinoluk:2015gia}, i.e.\ operators such
as
\be \label{eq:WilsonLine1Insertion}
{\cal U}_{[0,1]}^{i,ab}(x^+,y^+,y_\perp) = \int\limits_{y^+}^{x^+}dz^+ 
\frac{z^+-y^+}{x^+-y^+}\, {\cal U}^{ac}(x^+,z^+,y_\perp)\;
\left[igT^e_{cd} \partial_{y^i} A^{-,e}(z^+,y_\perp)\right]\;
{\cal U}^{db}(z^+,y^+,y_\perp)~,
\ee
where
\be
{\cal U}(x^+,y^+,y_\perp) = {\cal P}\, e^{ig\int\limits_{y^+}^{x^+}dz^+ 
T\cdot A^-(z^+,y_\perp)}
\ee
are the usual eikonal lines. Note that besides the electric field
insertion which is due to the finite target thickness the Wilson
line~(\ref{eq:WilsonLine1Insertion}) does run along the light
cone. For a study of kinematic corrections corresponding to Wilson
lines at a finite angle (rapidity) we refer to ref.~\cite{Babansky:2002my}.

The new Wilson lines with electric field insertions appear due to
quantum diffusion of the incident projectile in the transverse direction as
it passes through a target of finite thickness. At leading
order in the field of the target the above Wilson line simplifies to
\be
{\cal U}_{[0,1]}^{i,ab}(x^+,y^+,y_\perp) = \int\limits_{y^+}^{x^+}dz^+ 
\frac{z^+-y^+}{x^+-y^+}\,
\left[igT^e_{ab} \partial_{y^i} A^{-,e}(z^+,y_\perp)\right]
\ee
which suffices for the evaluation of the Lipatov vertex. The purpose
of this paper is to derive $L_i$ at next to next to eikonal (NNE)
accuracy; and to discuss the corrections to the single-inclusive gluon
production cross section at high transverse momentum at order
$\rho_T(k_1)\, \rho_T^*(k_2)$.

\begin{figure}[htb]
\begin{center}
\includegraphics[width=0.4\textwidth]{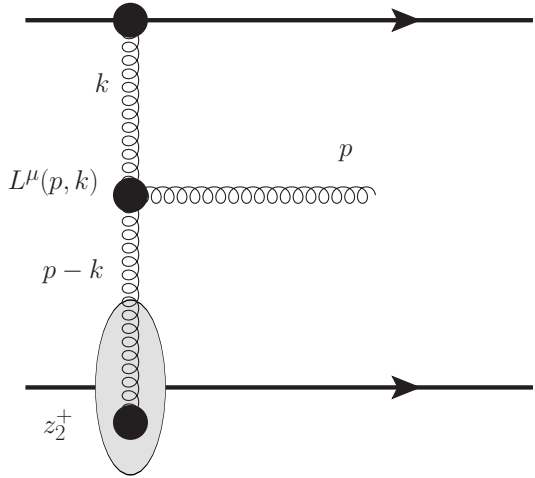}
\end{center}
\vspace*{-0.6cm}
\caption[a]{Fusion of the fields of two high-energy projectile and
  target charges described by the Lipatov vertex.}
\label{fig:L}
\end{figure}
Our result for the Lipatov vertex (in light-cone gauge $A^+=0$) at
NNE accuracy is
\bea
L^i(p,k) = - 2C^i(p,k) \; k^2 \left\{ 1   + \frac{i}{2} p^2
\frac{z_2^+}{p^+}  - \frac{1}{8}
\left(p^2 \frac{z_2^+}{p^+}\right)^2 \right\}~,  \label{eq:Li}
\eea
where
\be \label{eq:C}
C^i(p,k) = \frac{p^i}{p^2} - \frac{k^i}{k^2}~.
\ee
A derivation is given in appendix~\ref{app:L} and the corresponding
diagram is shown in fig.~\ref{fig:L}.  The first term in~(\ref{eq:Li})
corresponds to the eikonal (shock wave) limit while the second and
third terms are the NE and NNE corrections for a target of finite
thickness $\ell^+$, respectively. These corrections come with
additional factors of $z_2^+/p^+$ which is due to the above mentioned
quantum diffusion of the incident wave passing through the target. The
mean square deviation from the classical (eikonal) path is
proportional to $z_2^+/p^+$~\cite{Altinoluk:2015gia}.

The vertex from eq.~(\ref{eq:Li}) acts on a product of projectile and
target fields to generate the produced gluon radiation field in the
forward light cone,
\be \label{eq:M}
    {\cal M}^a_\lambda(p)  = \epsilon_\lambda^i\, p^2 A^{i,a}(p)~, 
\ee
with $p^2 A^{i,a}(p)$ as written in eq.~(\ref{eq:Field}) above.

To compute the single inclusive cross section we multiply
eq.~(\ref{eq:M}) with its complex conjugate, sum over gluon
polarizations and colors, and perform an average over the random color
charge densities of projectile and target. In the standard
McLerran-Venugopalan (MV) model~\cite{MV} this (target) average is
performed with the action
\be \label{eq:MVstd}
S_{\rm MV}[\rho] = \int d^2x_\perp \int\limits_0^{\ell^+}dx^+ \; \frac{{\rm tr}\,
  \rho(x^+,x_\perp) \rho(x^+,x_\perp)}{\mu^2}~,
\ee
which leads to the following color charge
correlator:
\be \label{eq:rhorho_local}
\left<
\rho^a(z_1^+,k_1) \, \rho^{*b}(z_2^+,k_2)
\right> = \delta^{ab} \, \delta(z_1^+ - z_2^+) \,
(2\pi)^2 \delta^2(k_1-k_2) \, \mu^2~.
\ee
$\mu^2$ denotes the mean color charge density (squared) per unit
transverse area and longitudinal phase space. Because color charge
correlations in the MV model are local in $z^+$, sub-eikonal
corrections, i.e.\ the curly brackets in eq.~(\ref{eq:Li}),
cancel\footnote{The most direct way to see this is to note that the
correction in curly brackets in eq.~(\ref{eq:Li}) corresponds to the first
three terms in the Taylor series expansion of the phase $\exp\left(
\frac{i}{2} p^2\frac{z_2^+}{p^+}\right)$.} in the
(absolute) square of the amplitude~(\ref{eq:M}).

One may also consider a generalization of the MV-model action where
the two color charge densities sit at different longitudinal
coordinates,
\be
S_{\rm eff}[\rho] = \int d^2x_\perp \int\limits_0^{\ell^+}dx^+
\int\limits_{x^+-\lambda^+}^{x^++\lambda^+}\frac{dy^+}{2\lambda^+} \frac{{\rm tr}\,
  \rho(x^+,x_\perp) U_{x^+\to y^+}\rho(y^+,x_\perp)U_{y^+\to x^+}}{\mu^2}~,
\ee
and are connected by gauge links along the longitudinal
axis. $\lambda^+$ denotes the color correlation length in the target
which should be on the order of the size of a nucleon.

At leading order in $gA^-$ we then consider the color charge
correlator
\be \label{eq:rhorho}
\left<
\rho^a(z_1^+,k_1) \, \rho^{*b}(z_2^+,k_2)
\right> = \delta^{ab} \, \Theta(\lambda^+ - |z_1^+ -
z_2^+|)\frac{1}{2\lambda^+} \, (2\pi)^2 \delta^2(k_1-k_2) \, \mu^2~.
\ee
This correlator reduces to the MV-model one from
eq.~(\ref{eq:rhorho_local}) in the limit $\lambda^+\to 0$.

In appendix~\ref{app:NNEsngl} we show that eq.~(\ref{eq:rhorho}) leads
to the single-inclusive cross section
\bea
p^+ \frac{d\sigma}{dp^+ d^2p\, d^2b} &=& 4 N_c (N_c^2-1) \, S_\perp
\frac{g^2}{p^2}
\left[1 - \frac{1}{6}\left(\frac{p^2\; \lambda^+}{2p^+}\right)^2\right]
\int \frac{d^2 k}{(2\pi)^2} \, \Phi_P(k^2) \, \Phi_T((p-k)^2)~.
  \label{eq:SingleIncl}
\eea
Here, $S_\perp$ denotes the transverse area of the collision.
In eq.~(\ref{eq:SingleIncl}) we introduced the unintegrated gluon
distribution of the target via
\be \label{eq:Phi}
\Phi_T(k^2) = g^2 \ell^+ \, \frac{\mu_T^2}{k^2}~,
\ee
and similar for the projectile.  This function is dimensionless and
proportional to the saturation momentum squared, $Q_s^2\sim g^4 \ell^+
\mu^2$. However, saturation of the gluon density at low $k^2$ is not
incorporated in~(\ref{eq:Phi}) which exhibits the perturbative $\sim
1/k^2$ growth down to low transverse momentum.

We note that in eq.~(\ref{eq:SingleIncl}) NE corrections $\sim
p^2\lambda^+/p^+$ drop out, see also
refs.~\cite{Altinoluk:2014oxa,Altinoluk:2015gia} and appendix
\ref{app:NNEsngl}. The NNE correction corresponds to the second term
in the square brackets. As already indicated above it is suppressed by
two powers of the light-cone momentum $p^+$ of the produced gluon but
increases with transverse momentum. Nevertheless, the NNE correction
to single-inclusive gluon production is seen not to exhibit nuclear
enhancement since it involves the color correlation scale $\lambda^+$
rather than the target thickness $\ell^+$. This occurs because the
gluon field of the target is evaluated at a longitudinal coordinate
$z_2^+$ in the amplitude, and $\bar{z}_2^+$ in the conjugate amplitude
which may differ at most by $\lambda^+$.

Next, we consider two gluon inclusive production. A detailed
derivation is provided in appendix~\ref{app:NNEdouble}. Here, we only
summarize the main results.

\begin{figure}[htb]
\begin{center}
\includegraphics[width=0.5\textwidth]{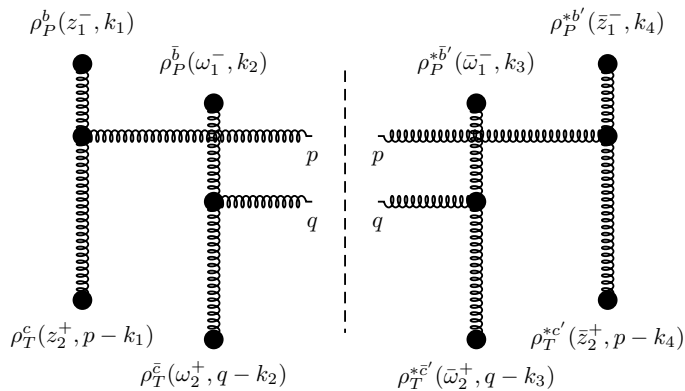}
\end{center}
\vspace*{-0.6cm}
\caption[a]{Double inclusive gluon production}
\label{fig:Double_inclsuive}
\end{figure}
In the linear approximation two gluon production corresponds to the
diagram shown in fig.~\ref{fig:Double_inclsuive} which has to be
summed over all possible contractions of the sources in the projectile
and target, respectively~\cite{Dumitru:2008wn}. Just as in the above
we assume that the projectile can be approximated by an infinitely
thin shock wave but we allow for a finite thickness $\ell^+\sim
A^{1/3}$ of the target. For simplicity, we restrict the discussion
here to the local (in $z^+$) model,
eqs.~(\ref{eq:MVstd},\ref{eq:rhorho_local}). In this case, sub-eikonal
corrections cancel when one performs a contraction of the same sources
in the amplitude and in its complex conjugate (denoted as ``type A''
diagrams in appendix~\ref{app:NNEdouble}). On the other hand, NNE
corrections do not cancel when $\rho_1$ is contracted with $\rho_2^*$
and $\rho_1^*$ is contracted with $\rho_2$ (``type B'' diagrams in
appendix~\ref{app:NNEdouble}), or when $\rho_1$ is contracted with
$\rho_2$ and $\rho_1^*$ with $\rho_2^*$ (``type C'' diagrams in
appendix~\ref{app:NNEdouble}). These diagrams again involve a correction
of order
\be
\left(\frac{\ell^+\, p^2}{p^+}\right)^2
\ee
which is proportional to $A^{2/3}$.
The full result for the inclusive two gluon production cross section
is
\bea
p^+q^+\frac{d\sigma}{dp^+d^2pdq^+d^2q}&=&
16N_c^2(N_c^2-1) \,
g^4\frac{S_{\perp}}{p^2q^2}\int
\frac{d^2k_1}{(2\pi)^2}\frac{d^2k_2}{(2\pi)^2}\Phi_P(k_1^2)\Phi_P(k_2^2)
\Phi_T\big[(p-k_1)^2\big]\Phi_T\big[(q-k_2)^2\big]\nonumber\\
& & \left[ \delta^{(2)}(k_1+k_2)
+ \delta^{(2)}(k_1-k_2) \right. \nonumber\\
&+&
\frac{1}{8}\delta^{(2)}(p-q-k_1+k_2)\left[1-\frac{1}{12}\left(\frac{p^2}{2p^+}-\frac{q^2}{2q^+}\right)^2{\ell^+}^2
\right]\nonumber\\
&&
\left.\times\left\{1+\frac{k_2^2(p-k_1)^2}{k_1^2(p+k_2)^2}-\frac{p^2(k_1+k_2)^2}{k_1^2(p+k_2)^2}\right\}
\left\{1+\frac{k_1^2(q-k_2)^2}{k_2^2(q+k_1)^2}-\frac{q^2(k_1+k_2)^2}{k_2^2(q+k_1)^2}\right\}
\right.\nonumber\\
&+&
\frac{1}{4}\delta^{(2)}(p-q)
\left[1-\frac{1}{12}\left(\frac{p^2}{2p^+}-\frac{q^2}{2q^+}\right)^2{\ell^+}^2\right]\nonumber\\
&&
\left.\times\left\{1+\frac{k_2^2(p-k_1)^2}{k_1^2(p-k_2)^2}-\frac{p^2(k_1-k_2)^2}{k_1^2(p-k_2)^2}\right\}
\left\{1+\frac{k_1^2(q-k_2)^2}{k_2^2(q-k_1)^2}-\frac{q^2(k_1-k_2)^2}{k_2^2(q-k_1)^2}\right\}\right.\nonumber\\
&+&
\delta^{(2)}(p-q-k_1+k_2)\left[1-\frac{1}{12}\left(\frac{p^2}{2p^+}-\frac{q^2}{2q^+}\right)^2{\ell^+}^2\right]
\nonumber\\
&+&
\frac{1}{4}\delta^{(2)}(p+q)\left[1-\frac{1}{12}\left(\frac{p^2}{2p^+}+\frac{q^2}{2q^+}\right)^2{\ell^+}^2\right]\nonumber\\
&&
\times
\left\{1+\frac{k_2^2(p-k_1)^2}{k_1^2(p+k_2)^2}-\frac{p^2(k_1+k_2)^2}{k_1^2(p+k_2)^2}\right\}
\left\{1+\frac{k_1^2(q-k_2)^2}{k_2^2(q+k_1)^2}-\frac{q^2(k_1+k_2)^2}{k_2^2(q+k_1)^2}\right\}\nonumber\\
&+&\frac{1}{8}\delta^{(2)}(p+q-k_1-k_2)\left[1-\frac{1}{12}\left(\frac{p^2}{2p^+}+\frac{q^2}{2q^+}\right)^2{\ell^+}^2\right]\nonumber\\
&&
\times
\left\{1+\frac{k_2^2(p-k_1)^2}{k_1^2(p-k_2)^2}-\frac{p^2(k_1-k_2)^2}{k_1^2(p-k_2)^2}\right\}
\left\{1+\frac{k_1^2(q-k_2)^2}{k_2^2(q-k_1)^2}-\frac{q^2(k_1-k_2)^2}{k_2^2(q-k_1)^2}\right\}\nonumber\\
&+&\delta^{(2)}(p+q-k_1-k_2)\left[1-\frac{1}{12}\left(\frac{p^2}{2p^+}+\frac{q^2}{2q^+}\right)^2{\ell^+}^2\right]\nonumber\\
& &
\Big]~.
\eea
This expression does not include the disconnected contribution
corresponding to uncorrelated production of the two gluons.
The fact that NNE corrections do appear in the two-gluon cross section
and that they are not the same for all diagrams could be important for
studies of two-particle azimuthal correlations. However, more detailed
computations with realistic unintegrated gluon densities, and
including the dijet contribution are
required~\cite{DuslingVenugopalan} for more quantitative statements.

In summary, in this paper we have evaluated explicitly Wilson lines
with electric field insertions to leading order in the field $gA^-$ of
the target. This determines next to eikonal (NE) and next to next to
eikonal (NNE) corrections to the Lipatov vertex which are proportional
to powers of the target thickness, and hence to $A^{1/3}$. From the
vertex we have derived a $k_T$-factorization formula valid up to NNE
accuracy. For single inclusive gluon production we find that
sub-eikonal corrections cancel if a model with local (in the
longitudinal coordinate $z^+$) color charge correlator for the target
is employed. On the other hand such corrections should be present in
models with color charge correlators with finite support. Furthermore,
NNE eikonal corrections also appear in correlated two gluon
production, even for a local target color charge correlator. Rather
than simply rescaling the two-gluon cross section we find that NNE
corrections depend on the type of contractions of the sources
in the target. Thus, such corrections may affect two-particle
angular correlations when $\ell^+\, p^2/p^+$ is not very small.

\vspace*{1cm}
\begin{acknowledgments}
T.A.\ expresses his gratitude to the Department of Natural Sciences of 
Baruch College for their warm hospitality during a visit when this
work was done. T.A.\ acknowledges
support by 
the People Programme (Marie Curie Actions) of the European Union's 
Seventh Framework Programme FP7/2007-2013/ under REA grant agreement \#318921; 
the European Research Council grant 
HotLHC ERC-2011-StG-279579, Ministerio de Ciencia e Innovaci\'on of 
Spain under project FPA2014-58293-C2-1-P, Xunta de Galicia 
(Conseller\'{\i}a de Educaci\'on and Conseller\'\i a de Innovaci\'on e
Industria - Programa Incite),  
the Spanish Consolider-Ingenio 2010 Programme CPAN and  FEDER.
A.D.\ gratefully acknowledges support by the DOE
Office of Nuclear Physics through Grant No.\ DE-FG02-09ER41620; and
from The City University of New York through the PSC-CUNY Research
grants 67119-0045 and 69362-0047.

\end{acknowledgments}

\appendix

\section{The Lipatov vertex to NNE level}  \label{app:L}

In this appendix we provide details of the calculation of NE and NNE
corrections to the Lipatov vertex. The gluon-nucleus reduced
amplitude\footnote{To obtain the field analogous to
  eq.~(\ref{eq:Field}) one would strip off the polarization vector,
  multiply by the projectile charge density $g\rho_P(z_1^-,k)$, and
  integrate over $dz_1^-$ and $d^2k/(2\pi)^2$.} at NNE accuracy
\cite{Altinoluk:2015gia} is given by
\bea
&&{\overline M}^{ab}_{\lambda}(\underline{p},k)= i\varepsilon^{*i}_{\lambda}
\int d^2x\, e^{ix\cdot(k-p)}\Bigg\{ 2C^i(p,k) \, \U(\ell^+,0;x)
+ \frac{\ell^+}{p^+}
\left[ \left( \delta^{ij}-2p^j\frac{k^i}{k^2}\right)\U^j_{[0,1]}(\ell^+,0;x)
-i\frac{k^i}{k^2}\U_{[1,0]}(\ell^+,0;x)
\right] \nonumber\\
&&~~~~~~~~~~~~~~~~~~~~~~
+
\left( \frac{\ell^+}{p^+}\right)^2 \Bigg[
-\frac{k^i}{k^2}p^jp^l\U^{jl}_{[0,2]}(\ell^+,0;x) -i\frac{k^i}{k^2}p^j
\U^j_{[1,1]}(\ell^+,0;x)
+\frac{1}{2}\frac{k^i}{k^2}\U_{[2,0]}(\ell^+,0;x)\nonumber\\
&&~~~~~~~~~~~~~~~~~~~~~~
+\frac{i}{4}\left( p^2\delta^{ij}-2p^ip^j\right)\U^j_{({\rm A})}(\ell^+,0;x)
+\frac{1}{4}p^j\U^{ij}_{({\rm B})}(\ell^+,0;x) + \frac{i}{4}
\U^{i}_{({\rm C})}(\ell^+,0;x)\Bigg]\Bigg\}\; ,
\label{eq:Aallorders}
\eea
where $(\underline p)\equiv(p^+,p)$\footnote{The expression written in
  ref.~\cite{Altinoluk:2015gia} is missing a factor of $1/2$ in the
  term $\sim \U^{jl}_{[0,2]}(\ell^+,0;x)$ which we have
  corrected.}. This expression is valid to all orders in the field of
the target. To compute the Lipatov vertex we expand the standard or
decorated Wilson lines to linear order in the charge density of the
target, $g\rho_T$. For the standard Wilson line,
\be
\int d^2x\, e^{ix\cdot(k-p)}\U(\ell^+,0;x)^{ab}=(2\pi)^2\delta(k-p)1^{ab} 
+ i\, g^2T^{ab}_c\frac{1}{(p-k)^2}
\int_0^{\ell^+}dz^+\rho_T^c(z^+,p-k)+O(\rho_T^2).
\ee
The decorated Wilson lines that appear at NE and NNE level are
Wilson lines with one or more background field insertions
along the longitudinal axis from $0$ to $\ell^+$. For the explicit
expressions of these decorated Wilson lines, we refer to
ref.~\cite{Altinoluk:2015gia}. The expansion of each field insertion
starts at linear order in $\rho_T$. Thus, terms with multiple
field insertions contribute at higher orders in $\rho_T$ and can
be neglected at linear order. Keeping this in mind, we obtain the
following expressions for the decorated Wilson lines:
\be
\int d^2x \,e^{ix\cdot(k-p)}\U^j_{[0,1]}(\ell^+,0,x)^{ab}=-g^2T^{ab}_c
\frac{(p-k)^j}{(p-k)^2}
\int_0^{\ell^+}dz^+\left(\frac{z^+}{\ell^+}\right)\rho^c_T(z^+,p-k)+O(\rho^2_T)
\ee
\be
\int d^2x\, e^{ix\cdot(k-p)}\U_{[1,0]}(\ell^+,0,x)^{ab}=-ig^2T^{ab}_c
\int_0^{\ell^+}dz^+\left(\frac{z^+}{\ell^+}\right)\rho^c_T(z^+,p-k)+O(\rho^2_T)
\ee
\be
\int d^2x\, e^{ix\cdot(k-p)}\U^{jl}_{[0,2]}(\ell^+,0,x)^{ab}=-ig^2T^{ab}_c
\frac{(p-k)^j(p-k)^l}{(p-k)^2}
\int_0^{\ell^+}dz^+\left(\frac{z^+}{\ell^+}\right)^2\rho^c_T(z^+,p-k)+O(\rho^2_T)
\ee
\be
\int d^2x\, e^{ix\cdot(k-p)}\U^{j}_{[1,1]}(\ell^+,0,x)^{ab}=g^2T^{ab}_c
(p-k)^j
\int_0^{\ell^+}dz^+\left(\frac{z^+}{\ell^+}\right)^2\rho^c_T(z^+,p-k)+O(\rho^2_T)
\ee
\bea
&&\int d^2x\, e^{ix\cdot(k-p)}\U_{[2,0]}(\ell^+,0,x)^{ab}=ig^2T^{ab}_c
(p-k)^2\frac{1}{2}
\int_0^{\ell^+}dz^+\left(\frac{z^+}{\ell^+}\right)^2\rho^c_T(z^+,p-k)+O(\rho^2_T) \\
&&\int d^2x\, e^{ix\cdot(k-p)}\U^j_{({\rm A})}(\ell^+,0,x)^{ab}=-g^2T^{ab}_c
\frac{(p-k)^j}{(p-k)^2}
\int_0^{\ell^+}dz^+\left(\frac{z^+}{\ell^+}\right)^2\rho^c_T(z^+,p-k)+O(\rho^2_T) \\
&&\int d^2x\, e^{ix\cdot(k-p)}\U^{ij}_{({\rm B})}(\ell^+,0,x)^{ab}=-ig^2T^{ab}_c
[\delta^{ij}\delta^{lm}+\delta^{il}\delta^{jm}+\delta^{im}\delta^{jl}]
\frac{(p-k)^l(p-k)^m}{(p-k)^2}\nonumber\\
&&~~~~~~~~~~~~~~~~~~~~~~~~~~~~~~~~~~~~~~~~~~
\times
\int_0^{\ell^+}dz^+\left(\frac{z^+}{\ell^+}\right)^2\rho^c_T(z^+,p-k)+O(\rho^2_T) \\
&&\int d^2x\, e^{ix\cdot(k-p)}\U^i_{({\rm C})}(\ell^+,0,x)^{ab}=g^2T^{ab}_c
(p-k)^i
\int_0^{\ell^+}dz^+\left(\frac{z^+}{\ell^+}\right)^2\rho^c_T(z^+,p-k)+O(\rho^2_T)
\eea
Using the expressions above, it is straightforward to obtain the
amplitude at order $\rho_T$ as
\be \label{eq:Avertex}
{\overline M}^{ab}_{\lambda}(\underline{p},k)=
i\varepsilon^{*i}_{\lambda}(ig^2)T^{ab}_c
\frac{1}{(p-k)^2} 
\int_0^{\ell^+}dz^+ ~ 2C^i(p,k) \Bigg\{ 1+\frac{i}{2}p^2\frac{z^+}{p^+}
-\frac{1}{8} \left(p^2 \frac{z^+}{p^+}\right)^2
\Bigg\} \rho_T^c(z^+,p-k)~.
\ee
One can now read off the Lipatov vertex at NNE accuracy as written in
eq.~(\ref{eq:Li}). The first term in the curly brackets corresponds to
a straight line trajectory. The second and third terms account for
corrections, at ${\cal O}(\ell^+)$ and ${\cal O}(\ell^{+ 2})$,
respectively, due to quantum diffusion from that classical path.  The
structure of the vertex in eq.~(\ref{eq:Avertex}) suggests that the
corrections to the amplitude due to a finite target thickness may
exponentiate to a phase,
\be
\Bigg\{ 1+\frac{i}{2}p^2\frac{z^+}{p^+}
-\frac{1}{8} \left(p^2 \frac{z^+}{p^+}\right)^2
\Bigg\} \rightarrow
\exp\left({\frac{i}{2}p^2\frac{z^+}{p^+}}\right)~.
\ee
However, a strict proof of exponentiation would require a
generalization of eq.~(\ref{eq:Aallorders}) to all orders in
$\ell^+/p^+$.

\section{Single inclusive gluon production at NNE accuracy}  \label{app:NNEsngl}
The single inclusive gluon production cross section is given by
\bea
& & f^{abc} f^{ab'c'} g^6 \int \frac{d^2k_1}{(2\pi)^2}\frac{d^2k_2}{(2\pi)^2}
\int dz_1^- dz_2^+d\bar z_1^- d\bar z_2^+
\frac{L_i(p,k_1) L^*_i(p,k_2)}{k_1^2 k_2^2 (p-k_1)^2 (p-k_2)^2}
\nonumber\\
& &~~~~~~~~~~~~~~\times
\left<\rho^b(z_1^-,k_1)\, \rho^{*b'}(\bar z_1^-,k_2)\right>_P \;
\left<\rho^c(z_2^+,p-k_1)\, \rho^{*c'}(\bar z_2^+,p-k_2)\right>_T
~.
\eea
With the (re-exponentiated) Lipatov vertex from above and the color charge
correlator from eq.~(\ref{eq:rhorho}) this becomes
\bea
& & 4 N_c (N_c^2-1) \, g^4 S_\perp \int \frac{d^2k}{(2\pi)^2}
\frac{g^2\int dz_1^- \mu^2_P}{k^2}\;
\frac{k^2}{(p-k)^4} \mu^2_T\;
C^i(p,k) C^i(p,k)  \nonumber\\
& & ~~~~~~~\times \int \frac{dz_2^+ 
d\bar{z}_2^+}{2\lambda^+}\, \Theta(\lambda^+-|z_2^+-\bar{z}_2^+|) \, e^{ip^2(z_2^+-\bar{z}_2^+)/2p^+}\\
&=& 4 N_c (N_c^2-1) \, g^2 \frac{S_\perp}{p^2}
\frac{2p^+}{p^2\, \lambda^+}
\sin\left(\frac{p^2\, \lambda^+}{2p^+}\right)
\int\frac{d^2k}{(2\pi)^2} \Phi_P(k^2)
\frac{g^2 \ell^+\, \mu_T^2}{(p-k)^2}~.  \label{eq:appSingleIncl1}
\eea
We have assumed that $\lambda^+\ll\ell^+$.
Substituting the unintegrated gluon distribution of the target
introduced in eq.~(\ref{eq:Phi}) and expanding to second order in
$\lambda^+$ finally leads to
\be
4 N_c (N_c^2-1) \, g^2 \frac{S_\perp}{p^2}
\left[1-\frac{1}{6}\left(\frac{p^2\, \lambda^+}{2p^+}\right)^2\right]
\int\frac{d^2k}{(2\pi)^2} \Phi_P(k^2)\,\Phi_T((p-k)^2)~.
\ee
%

\section{Double inclusive gluon production at NNE accuracy} 
\label{app:NNEdouble}
The inclusive two gluon production cross section is given by 
\bea
p^+q^+\frac{d\sigma}{dp^+d^2pdq^+d^2q}&=& f^{abc}f^{\bar{a}\bar{b}\bar{c}}f^{\bar{a}\bar{b}'\bar{c}'}f^{ab'c'}g^{12}
\int \frac{d^2k_1}{(2\pi)^2} \frac{d^2k_2}{(2\pi)^2} \frac{d^2k_3}{(2\pi)^2} \frac{d^2k_4}{(2\pi)^2}
\int dz^-_1d\bar{z}^-_1d\omega^-_1d\bar{\omega}^-_1
dz^+_2d\bar{z}^+_2d\omega^+_2d\bar{\omega}^+_2\nonumber\\
&&
\hspace{-2cm}
\times \frac{L^i(p,k_1;z^+_2)}{k_1^2(p-k_1)^2} \frac{L^{*i}(p,k_4;\bar{z}^+_2)}{k_4^2(p-k_4)^2} \frac{L^j(q,k_2;\omega^+_2)}{k_2^2(q-k_2)^2} \frac{L^{*j}(q,k_3;\bar{\omega}^+_2)}{k_3^2(q-k_3)^2}
\left\langle\rho^b(z^-_1,k_1) \rho^{\bar{b}}(\omega^-_1,k_2) \rho^{*\bar{b}'}(\bar{\omega}^-_1,k_3) \rho^{*b'}(\bar{z}^-_1,k_4)  
\right\rangle _{P}\nonumber\\
&&
\times
\left\langle\rho^c(z^+_2,p-k_1) \rho^{\bar{c}}(\omega^+_2,q-k_2) \rho^{*\bar{c}'}(\bar{\omega}^+_2,q-k_3) \rho^{*c'}(\bar{z}^+_2,p-k_4)  
\right\rangle _{T}
\label{eq:double_inclusive}
\eea
We shall use the local correlator of color charges as written in
eq.~(\ref{eq:rhorho_local}).

Type A contributions correspond to the following contraction on the
target side:
\bea
\left\langle\rho^c(z^+_2,p-k_1) \rho^{\bar{c}}(\omega^+_2,q-k_2) \rho^{*\bar{c}'}(\bar{\omega}^+_2,q-k_3) \rho^{*c'}(\bar{z}^+_2,p-k_4)  
\right\rangle _{T}&\to&\left\langle\rho^c(z^+_2,p-k_1)
\rho^{*c'}(\bar{z}^+_2,p-k_4)
\right\rangle _{T}\nonumber\\
&&\hspace{-0.5cm}
\times
\left\langle \rho^{\bar{c}}(\omega^+_2,q-k_2) \rho^{*\bar{c}'}(\bar{\omega}^+_2,q-k_3) \right\rangle _{T}\; .
\label{eq:TypeA_contractions}
\eea
Using eq.~(\ref{eq:rhorho_local}) it is straightforward to see that
Type A contributions to the cross section are proportional to
\begin{equation}
(2\pi)^4 \delta^{cc'}
\delta^{\bar{c}\bar{c}'}\delta(z_2^+-\bar{z}_2^+)\delta(\omega_2^+-\bar{\omega}_2^+)\delta^{(2)}(k_1-k_4)\delta^{(2)}(k_2-k_3)\mu^2_T({z_2}^+)\mu^2_T(\omega_2^+)\; .
\label{eq:TypeA_delta}
\end{equation} 
However, realizing the longitudinal $\delta$-functions in
eq.~(\ref{eq:TypeA_delta}), one can easily see that sub-eikonal
corrections to the Type A contributions for the double inclusive gluon
production cross section vanish.

\begin{figure}[htb]
\begin{center}
\includegraphics[width=0.9\textwidth]{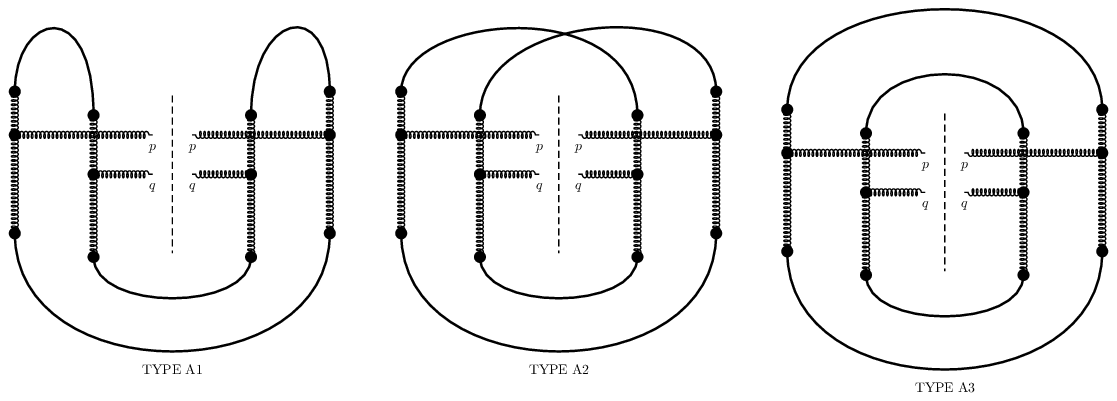}
\end{center}
\vspace*{-0.6cm}
\caption[a]{Type A contributions to the double inclusive gluon production}
\label{fig:TypeA}
\end{figure}
After performing the color contractions on the projectile side we get
three types of diagrams: Type A1, Type A2 and Type A3 (see
Fig.~\ref{fig:TypeA}):
\bea {\rm Type \, A1}\propto
(2\pi)^2\delta^{b\bar{b}}\delta^{\bar{b}'b'}\delta(z_1^- -
\omega_1^-)\delta(\bar{z}_1^- -
\bar{\omega}_1^-)\delta^{(2)}(k_1+k_2)\delta^{(2)}(k_3+k_4)\mu^2_P(z_1^-)\mu^2_P(\bar{z}_1^-)\;
\label{eq:Type1} \\
{\rm Type \, A2}\propto (2\pi)^2\delta^{b\bar{b}'}\delta^{\bar{b}b'}\delta(z_1^- - \bar{\omega}_1^-)\delta(\bar{z}_1^- - \omega_1^-)\delta^{(2)}(k_1-k_3)\delta^{(2)}(k_2-k_4)\mu^2_P(z_1^-)\mu^2_P(\bar{z}_1^-)\; 
\label{eq:Type2} \\
{\rm Type \, A3}\propto (2\pi)^2\delta^{bb'}\delta^{\bar{b}\bar{b}'}\delta(z_1^- - \bar{z}_1^-)\delta(\omega_1^- -\bar{\omega}_1^-)\delta^{(2)}(k_1-k_4)\delta^{(2)}(k_2-k_3)\mu^2_P(z_1^-)\mu^2_P(\bar{\omega}_1^-)\; .
\label{eq:Type3}
\eea
Using eqs.~(\ref{eq:TypeA_delta} - \ref{eq:Type3}) these contributions
become
\bea
{\rm Type\,  A1}= f^{abc}f^{\bar{a}b\bar{c}}f^{\bar{a}\bar{b}\bar{c}}f^{a\bar{b}c}S_{\perp}g^8\int \frac{d^2k_1}{(2\pi)^2}\frac{d^2k_3}{(2\pi)^2}\delta^{(2)}(k_1+k_3)\Phi_P(k_1^2)\Phi_P(k_3^2)\int dz^+_2 d\omega^+_2 \mu^2_T(z_2^+)\mu^2_T(\omega_2^+)\nonumber\\
2^4C^i(p,k_1)C^i(p,-k_3)C^j(q,-k_1)C^j(q,k_3)\frac{k_1^2k_3^2}{(p-k_1)^2(p+k_3)^2(q+k_1)^2(q-k_3)^2}\\
{\rm Type\,  A2}= f^{abc}f^{\bar{a}\bar{b}\bar{c}}f^{\bar{a}b\bar{c}}f^{a\bar{b}c}S_{\perp}g^8\int \frac{d^2k_1}{(2\pi)^2}\frac{d^2k_2}{(2\pi)^2}\delta^{(2)}(k_1-k_2)\Phi_P(k_1^2)\Phi_P(k_2^2)\int dz^+_2 d\omega^+_2 \mu^2_T(z_2^+)\mu^2_T(\omega_2^+)\nonumber\\
2^4C^i(p,k_1)C^i(p,k_1)C^j(q,k_2)C^j(q,k_2)\frac{k_1^2k_2^2}{(p-k_1)^4(q-k_2)^4}\\
{\rm Type\,  A3}= f^{abc}f^{\bar{a}\bar{b}\bar{c}}f^{\bar{a}\bar{b}\bar{c}}f^{abc}S^2_{\perp}g^8\int \frac{d^2k_1}{(2\pi)^2}\frac{d^2k_2}{(2\pi)^2}\Phi_P(k_1^2)\Phi_P(k_2^2)\int dz^+_2 d\omega^+_2 \mu^2_T(z_2^+)\mu^2_T(\omega_2^+)\nonumber\\
2^4C^i(p,k_1)C^i(p,k_1)C^j(q,k_2)C^j(q,k_2)\frac{k_1^2k_2^2}{(p-k_1)^4(q-k_2)^4}
\eea
Note that we have used eq.~(\ref{eq:Phi}) to define the unintegrated
gluon distribution of the projectile and eq.~(\ref{eq:Li}) for the
definition of the Lipatov vertex at NNE accuracy.  To simplify the
expressions further we integrate over the longitudinal coordinate and
substitute the unintegrated gluon distribution of the target:
\bea {\rm Type\, A1}&=&16N_c^2(N_c^2-1) \,
g^4\frac{S_{\perp}}{p^2q^2}\int
\frac{d^2k_1}{(2\pi)^2}\frac{d^2k_2}{(2\pi)^2}\delta^{(2)}(k_1+k_2)
\Phi_P(k_1^2)\Phi_T\big[(p-k_1)^2\big]\Phi_P(k_2^2)\Phi_T\big[(q-k_2)^2\big]\\ {\rm
  Type\, A2}&=&16N_c^2(N_c^2-1) \, g^4\frac{S_{\perp}}{p^2q^2}\int
\frac{d^2k_1}{(2\pi)^2}\frac{d^2k_2}{(2\pi)^2}\delta^{(2)}(k_1-k_2)
\Phi_P(k_1^2)\Phi_T\big[(p-k_1)^2\big]\Phi_P(k_2^2)\Phi_T\big[(q-k_2)^2\big]\\ {\rm
  Type\, A3}&=&16N_c^2(N_c^2-1)^2 \,
g^4\frac{S^2_{\perp}}{p^2q^2}\int
\frac{d^2k_1}{(2\pi)^2}\frac{d^2k_2}{(2\pi)^2}
\Phi_P(k_1^2)\Phi_T\big[(p-k_1)^2\big]\Phi_P(k_2^2)\Phi_T\big[(q-k_2)^2\big]~. 
\label{eq:AppC12}
\eea
As already mentioned above subeikonal corrections vanish for these diagrams.
The last expression, eq.~(\ref{eq:AppC12}), is of course nothing but
the uncorrelated square of the single inclusive cross section.

\begin{figure}[htb]
\begin{center}
\includegraphics[width=0.9\textwidth]{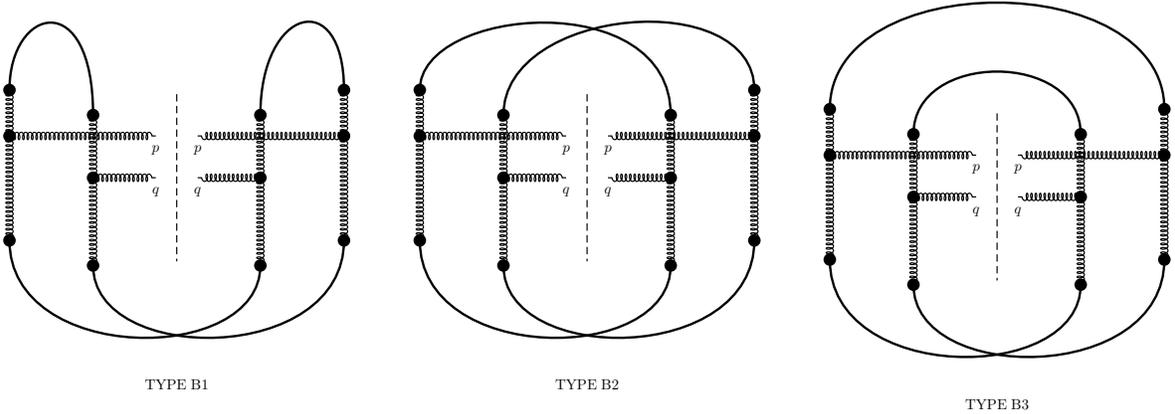}
\end{center}
\vspace*{-0.6cm}
\caption[a]{Type B contributions to the double inclusive gluon production}
\label{fig:TypeB}
\end{figure}
Type B contributions correspond to the following color contraction on
the target side:
\bea
\left\langle\rho^c(z^+_2,p-k_1) \rho^{\bar{c}}(\omega^+_2,q-k_2) \rho^{*\bar{c}'}(\bar{\omega}^+_2,q-k_3) \rho^{*c'}(\bar{z}^+_2,p-k_4)  
\right\rangle _{T}&\to&\left\langle\rho^c(z^+_2,p-k_1) \rho^{*\bar{c}'}(\bar{\omega}^+_2,q-k_3)  
\right\rangle _{T}\nonumber\\
&&\hspace{-0.5cm}
\times
\left\langle \rho^{\bar{c}}(\omega^+_2,q-k_2) \rho^{*c'}(\bar{z}^+_2,p-k_4) \right\rangle _{T}\; .
\label{eq:TypeB_contractions}
\eea
Again, by using eq.~(\ref{eq:rhorho_local}), one finds that Type
B contributions are proportional to
\be
(2\pi)^4\delta^{c\bar{c}'}\delta^{\bar{c}c'}\delta(z_2^+-\bar{\omega}_2^+)\delta(\omega_2^+-\bar{z}_2^+)\delta^{(2)}(p-q+k_3-k_1)\delta^{(2)}(q-p+k_4-k_2)\mu^2_T(z_2^+)\mu^2_T(\omega_2^+)\;.
\ee
The color contractions on the projectile side are the same as Type A
diagrams. Hence, one can immediately write the Type B contributions to
the double inclusive gluon production cross section as
\bea
{\rm Type\;  B1}&=&
f^{abc}f^{\bar{a}b\bar{c}}f^{\bar{a}\bar{b}c}f^{a\bar{b}\bar{c}}S_{\perp}g^8\int
\frac{d^2k_1}{(2\pi)^2}\frac{d^2k_3}{(2\pi)^2}\delta^{(2)}(p-q-k_1+k_3)\Phi_P(k_1^2)\Phi_P(k_3^2)\int
dz^+_2 d\bar{z}^+_2 \mu^2_T(z_2^+)\mu^2_T(\bar{z}_2^+) 2^4\nonumber\\
& & \hspace{-2cm}
C^i(p,k_1)C^i(p,-k_3)C^j(q,-k_1)C^j(q,k_3)\frac{k_1^2k_3^2}{(p-k_1)^2(p+k_3)^2(q+k_1)^2(q-k_3)^2}\left\{
1-\frac{1}{8}\left(\frac{p^2}{p^+}-\frac{q^2}{q^+}\right)^2(z^+_2-\bar{z}_2^+)^2\right\}\nonumber\\
&=&2N_c^2(N_c^2-1)g^4\frac{S_{\perp}}{p^2q^2}\int \frac{d^2k_1}{(2\pi)^2}\frac{d^2k_2}{(2\pi)^2}\delta^{(2)}(p-q-k_1+k_2)
\Phi_P(k_1^2)\Phi_T\big[(p-k_1)^2\big]\Phi_P(k_2^2)\Phi_T\big[(q-k_2)^2\big]\nonumber\\
&&
\hspace{-1.5cm}
\times\left\{1+\frac{k_2^2(p-k_1)^2}{k_1^2(p+k_2)^2}-\frac{p^2(k_1+k_2)^2}{k_1^2(p+k_2)^2}\right\}
\left\{1+\frac{k_1^2(q-k_2)^2}{k_2^2(q+k_1)^2}-\frac{q^2(k_1+k_2)^2}{k_2^2(q+k_1)^2}\right\}
\left[1-\frac{1}{12}\left(\frac{p^2}{2p^+}-\frac{q^2}{2q^+}\right)^2{\ell^+}^2
\right]\\
{\rm Type\;  B2}&=& f^{abc}f^{\bar{a}\bar{b}\bar{c}}f^{\bar{a}bc}f^{a\bar{b}\bar{c}}S_{\perp}g^8 \delta^{(2)}(p-q) \int \frac{d^2k_1}{(2\pi)^2}\frac{d^2k_2}{(2\pi)^2}\Phi_P(k_1^2)\Phi_P(k_2^2)\int dz^+_2 d\bar{z}^+_2 \mu^2_T(z_2^+)\mu^2_T(\bar{z}_2^+) 2^4\nonumber\\
& & C^i(p,k_1)C^i(p,k_2)C^j(q,k_1)C^j(q,k_2)\frac{k_1^2k_2^2}{(p-k_1)^2(p-k_2)^2(q-k_1)^2(q-k_2)^2}\left\{
1-\frac{1}{8}\left(\frac{p^2}{p^+}-\frac{q^2}{q^+}\right)^2(z^+_2-\bar{z}_2^+)^2\right\}\nonumber\\
&=&4N_c^2(N_c^2-1)g^4\frac{S_{\perp}}{p^2q^2}\delta^{(2)}(p-q)\int \frac{d^2k_1}{(2\pi)^2}\frac{d^2k_2}{(2\pi)^2}
\Phi_P(k_1^2)\Phi_T\big[(p-k_1)^2\big]\Phi_P(k_2^2)\Phi_T\big[(q-k_2)^2\big]\nonumber\\
&&
\hspace{-1.5cm}
\times\left\{1+\frac{k_2^2(p-k_1)^2}{k_1^2(p-k_2)^2}-\frac{p^2(k_1-k_2)^2}{k_1^2(p-k_2)^2}\right\}
\left\{1+\frac{k_1^2(q-k_2)^2}{k_2^2(q-k_1)^2}-\frac{q^2(k_1-k_2)^2}{k_2^2(q-k_1)^2}\right\}
\left[1-\frac{1}{12}\left(\frac{p^2}{2p^+}-\frac{q^2}{2q^+}\right)^2{\ell^+}^2\right] \\
{\rm Type\; B3}&=& f^{abc}f^{\bar{a}\bar{b}\bar{c}}f^{\bar{a}\bar{b}c}f^{ab\bar{c}}S_{\perp}g^8\int \frac{d^2k_1}{(2\pi)^2}\frac{d^2k_2}{(2\pi)^2}\delta^{(2)}(p-q-k_1+k_2)\Phi_P(k_1^2)\Phi_P(k_2^2)\int dz^+_2 d\bar{z}^+_2 \mu^2_T(z_2^+)\mu^2_T(\bar{z}_2^+) 2^4\nonumber\\
& &
C^i(p,k_1)C^i(p,k_1)C^j(q,k_2)C^j(q,k_2)\frac{k_1^2k_2^2}{(p-k_1)^4(q-k_2)^4}\left\{
1-\frac{1}{8}\left(\frac{p^2}{p^+}-\frac{q^2}{q^+}\right)^2(z^+_2-\bar{z}_2^+)^2\right\}
\nonumber\\
&=&16N_c^2(N_c^2-1)g^4\frac{S_{\perp}}{p^2q^2}
\int \frac{d^2k_1}{(2\pi)^2}\frac{d^2k_2}{(2\pi)^2}\delta^{(2)}(p-q-k_1+k_2)
\Phi_P(k_1^2)\Phi_T\big[(p-k_1)^2\big]\Phi_P(k_2^2)\Phi_T\big[(q-k_2)^2\big]\nonumber\\
&&\times
\left[1-\frac{1}{12}\left(\frac{p^2}{2p^+}-\frac{q^2}{2q^+}\right)^2{\ell^+}^2\right]
\eea   
To perform the longitudinal integrations explicitly we have assumed
that $\mu^2_T$ is constant.  Note that for the Type B contributions,
next to eikonal contributions to the cross section vanish due to
integration over $z_2^+$ and $\bar{z}^+_2$. However, the next to next
to eikonal corrections to the cross sections do not vanish.  For Type
B1 and Type B2 contributions we have used
\bea
C^i(p,k_1)C^i(p,k_2)&=&\left(\frac{p^i}{p^2}-\frac{k_1^i}{k_1^2}\right)\left(\frac{p^i}{p^2}-\frac{k_2^i}{k_2^2}\right)=\frac{1}{2p^2k_1^2k_2^2}\left[k_1^2(p-k_2)^2+k_2^2(p-k_1)^2-p^2(k_1-k_2)^2\right]
\eea 

\begin{figure}[htb]
\begin{center}
\includegraphics[width=0.9\textwidth]{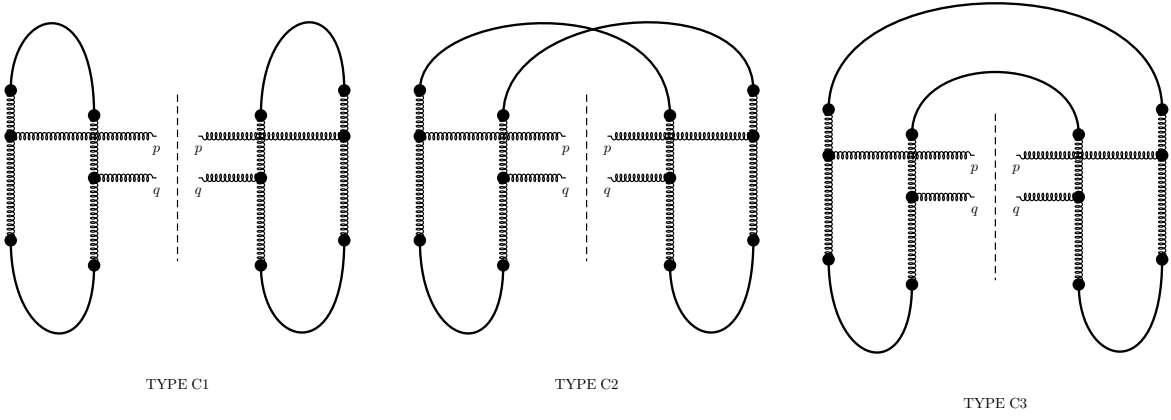}
\end{center}
\vspace*{-0.6cm}
\caption[a]{Type C contributions to the double inclusive gluon production}
\label{fig:TypeC}
\end{figure}
The color contractions on the target side for Type C contributions are given by 
\bea
\left\langle\rho^c(z^+_2,p-k_1) \rho^{\bar{c}}(\omega^+_2,q-k_2) \rho^{*\bar{c}'}(\bar{\omega}^+_2,q-k_3) \rho^{*c'}(\bar{z}^+_2,p-k_4)  
\right\rangle _{T}&\to&\left\langle\rho^c(z^+_2,p-k_1) \rho^{*\bar{c}}(\omega^+_2,q-k_2)  
\right\rangle _{T}\nonumber\\
&&\hspace{-0.5cm}
\times
\left\langle \rho^{\bar{c}'}(\bar{\omega}^+_2,q-k_3) \rho^{*c'}(\bar{z}^+_2,p-k_4) \right\rangle _{T}\; .
\label{eq:TypeA_contractionsII}
\eea
Thus, they are proportional to
\be
(2\pi)^4\delta^{c\bar{c}}\delta^{\bar{c}'c'}\delta(z_2^+-\omega^+_2)\delta(\bar{z}_2^+-\bar{\omega}_2^+)\delta^{(2)}(p+q-k_1-k_2)\delta^{(2)}(p+q-k_3-k_4)\mu^2_T(z_2^+)\mu^2_T(\bar{z}_2^+) \; .
\ee
Since the color contractions on the projectile side are the same as
Type A and Type B diagrams, one can write the Type C contributions to
the double inclusive gluon production cross section as follows:
\bea
{\rm Type\;  C1}&=& f^{abc}f^{\bar{a}bc}f^{\bar{a}\bar{b}\bar{c}}f^{a\bar{b}\bar{c}}S_{\perp}g^8 \delta^{(2)}(p+q) \int \frac{d^2k_1}{(2\pi)^2}\frac{d^2k_3}{(2\pi)^2}\Phi_P(k_1^2)\Phi_P(k_3^2)\int dz^+_2 d\bar{z}^+_2 \mu^2_T(z_2^+)\mu^2_T(\bar{z}_2^+) 2^4\nonumber\\
& &
C^i(p,k_1)C^i(p,-k_3)C^j(q,-k_1)C^j(q,k_3)\frac{k_1^2k_3^2}{(p-k_1)^2(p+k_3)^2(q+k_1)^2(q-k_3)^2}\left\{
1-\frac{1}{8}\left(\frac{p^2}{p^+}+\frac{q^2}{q^+}\right)^2(z^+_2-\bar{z}_2^+)^2\right\}\nonumber\\
&=&4N_c^2(N_c^2-1)g^4\frac{S_{\perp}}{p^2q^2}\delta^{(2)}(p+q)\int \frac{d^2k_1}{(2\pi)^2}\frac{d^2k_2}{(2\pi)^2}
\Phi_P(k_1^2)\Phi_T\big[(p-k_1)^2\big]\Phi_P(k_2^2)\Phi_T\big[(q-k_2)^2\big]\nonumber\\
&&\hspace{-1.5cm}
\times\left\{1+\frac{k_2^2(p-k_1)^2}{k_1^2(p+k_2)^2}-\frac{p^2(k_1+k_2)^2}{k_1^2(p+k_2)^2}\right\}
\left\{1+\frac{k_1^2(q-k_2)^2}{k_2^2(q+k_1)^2}-\frac{q^2(k_1+k_2)^2}{k_2^2(q+k_1)^2}\right\}
\left[1-\frac{1}{12}\left(\frac{p^2}{2p^+}+\frac{q^2}{2q^+}\right)^2{\ell^+}^2\right]\\
{\rm Type\;  C2}&=& f^{abc}f^{\bar{a}\bar{b}c}f^{\bar{a}b\bar{c}}f^{a\bar{b}\bar{c}}S_{\perp}g^8\int \frac{d^2k_1}{(2\pi)^2}\frac{d^2k_2}{(2\pi)^2}\delta^{(2)}(p+q-k_1-k_2)\Phi_P(k_1^2)\Phi_P(k_2^2)\int dz^+_2 d\bar{z}^+_2 \mu^2_T(z_2^+)\mu^2_T(\bar{z}_2^+) 2^4\nonumber\\
&&
C^i(p,k_1)C^i(p,k_2)C^j(q,k_1)C^j(q,k_2)\frac{k_1^2k_2^2}{(p-k_1)^2(p-k_2)^2(q-k_1)^2(q-k_2)^2}\left\{
1-\frac{1}{8}\left(\frac{p^2}{p^+}+\frac{q^2}{q^+}\right)^2(z^+_2-\bar{z}_2^+)^2\right\}\nonumber\\
&=&2N_c^2(N_c^2-1)g^4\frac{S_{\perp}}{p^2q^2}\int \frac{d^2k_1}{(2\pi)^2}\frac{d^2k_2}{(2\pi)^2}\delta^{(2)}(p+q-k_1-k_2)
\Phi_P(k_1^2)\Phi_T\big[(p-k_1)^2\big]\Phi_P(k_2^2)\Phi_T\big[(q-k_2)^2\big]\nonumber\\
&&\hspace{-1.5cm}
\times\left\{1+\frac{k_2^2(p-k_1)^2}{k_1^2(p-k_2)^2}-\frac{p^2(k_1-k_2)^2}{k_1^2(p-k_2)^2}\right\}
\left\{1+\frac{k_1^2(q-k_2)^2}{k_2^2(q-k_1)^2}-\frac{q^2(k_1-k_2)^2}{k_2^2(q-k_1)^2}\right\}
\left[1-\frac{1}{12}\left(\frac{p^2}{2p^+}+\frac{q^2}{2q^+}\right)^2{\ell^+}^2\right]\\
{\rm Type\; C3}&=& f^{abc}f^{\bar{a}\bar{b}c}f^{\bar{a}\bar{b}\bar{c}}f^{ab\bar{c}}S_{\perp}g^8\int \frac{d^2k_1}{(2\pi)^2}\frac{d^2k_2}{(2\pi)^2}\delta^{(2)}(p+q-k_1-k_2)\Phi_P(k_1^2)\Phi_P(k_2^2)\int dz^+_2 d\bar{z}^+_2 \mu^2_T(z_2^+)\mu^2_T(\bar{z}_2^+) 2^4\nonumber\\
&&
C^i(p,k_1)C^i(p,k_1)C^j(q,k_2)C^j(q,k_2)\frac{k_1^2k_2^2}{(p-k_1)^4(q-k_2)^4}\left\{ 1-\frac{1}{8}\left(\frac{p^2}{p^+}+\frac{q^2}{q^+}\right)^2(z^+_2-\bar{z}_2^+)^2\right\}\nonumber\\
&=&16N_c^2(N_c^2-1)g^4\frac{S_{\perp}}{p^2q^2}\int \frac{d^2k_1}{(2\pi)^2}\frac{d^2k_2}{(2\pi)^2}\delta^{(2)}(p+q-k_1-k_2)
\Phi_P(k_1^2)\Phi_T\big[(p-k_1)^2\big]\Phi_P(k_2^2)\Phi_T\big[(q-k_2)^2\big]\nonumber\\
&&
\times\left[1-\frac{1}{12}\left(\frac{p^2}{2p^+}+\frac{q^2}{2q^+}\right)^2{\ell^+}^2\right]~.
\eea


\end{document}